\documentclass[10pt]{article}
\newcommand{\add}[1]{{\color{black} {#1}}}
\usepackage{amsmath}
\usepackage{amssymb}

\usepackage{graphicx}
\usepackage{epsfig} 
\usepackage{cite}

\usepackage{color} 

\usepackage[labelfont=bf,labelsep=period,justification=raggedright]{caption}

\makeatletter
\renewcommand{\@biblabel}[1]{\quad#1.}
\makeatother

\date{}

\begin{document}

\begin{flushleft}
{\Large
\textbf{Yielding and irreversible deformation below the microscale: Surface effects and non-mean-field plastic avalanches}
}
\\
Paolo Moretti$^{1,2,\ast}$, 
Benedetta Cerruti$^{1,3}$, 
M.-Carmen Miguel$^{1}$
\\
\bf{1} Departament de F\'{\i}sica Fonamental, Facultat de F\'{\i}sica, Universitat de Barcelona, Barcelona, Spain
\\
\bf{2} Departament de F\'\i sica i Enginyeria Nuclear, Universitat Polit\` ecnica de Catalunya, Barcelona, Spain
\\
\bf{3} Institute for Cancer Research and Treatment, Candiolo, Turin, Italy
\\
$\ast$ E-mail: paolo.moretti@upc.edu
\end{flushleft}

\section*{Abstract}
Nanoindentation techniques recently developed to measure the mechanical response of crystals under external loading conditions reveal new phenomena upon decreasing sample size below the microscale. At small length scales, material resistance to irreversible deformation depends on sample morphology. Here we study the mechanisms of yield and plastic flow in inherently {\it small} crystals under uniaxial compression. Discrete structural rearrangements  emerge as  series of abrupt discontinuities in stress-strain curves. We obtain the theoretical dependence of the yield stress on system size and geometry and elucidate the statistical properties of plastic deformation at such scales.  Our results show that the absence of dislocation storage leads to crucial effects on the statistics of plastic events, ultimately affecting the {\it universal} scaling behavior observed at larger scales. 


\section*{Introduction}
Over the past years, experimental investigations have gathered increasing evidence that plastic deformation of crystalline materials proceeds through intermittent bursts of activity \cite{WEI-97,MIG-01,WEI-03,UCH-04,RIC-05,DIM-06,RIC-06,SCH-06}. Plastic flow advances through a sequence of strain avalanches of broadly distributed sizes. Avalanches have been observed experimentally under stress control conditions at various scales. Although such plastic fluctuations may be hard to detect at macroscopic scales, they dramatically affect mechanical properties of crystalline materials at smaller scales, ultimately producing detrimental effects on material formability \cite{CSI-07}. The introduction of microcrystal compression testing \cite{UCH-04} has allowed access to the microscale and introduced sample size as a crucial variable in this scenario. Size effects have dramatic consequences on yield, making smaller samples harder to deform and more unpredictable \cite{DIM-06}. It is then natural to wonder if such behavior would hold unchanged below the micrometer scale. 

While several nanoindentation techniques have been developed to measure material resistance to irreversible deformation and plastic flow, the experimental observation of plasticity at the nanoscale still represents an enormous challenge \cite{Cook-10}. Colloidal crystals, however, were proven to deform plastically by activating dislocation motion, in remarkable analogy with crystalline materials \cite{SCH-04,PER-05,SCH-nature, Suresh-06}. The unmatched advantage of working with such systems is brought in by their size. Micrometer-sized colloidal crystals may contain few thousands, or even hundreds, of particles. This aspect allows one to project the problem of nanomechanics into a length scale that is easily accessible on experimental grounds. In this work, we address some of the challenges posed by nanomechanics, by providing a theoretical study of the mechanisms of yield and plastic flow in inherently {\it small} systems. We propose a simple geometry, which can be reproduced in experiments on two-dimensional micrometer colloidal crystals, and provide robust input for mechanical testing of crystalline thin films below the micrometer scale.  Our aim is to show that at very small scales the irreversible deformation of materials proceeds in a novel way, deviating from the allegedly universal behavior observed in larger systems. To this end, we investigate the dependence of the yield stress on system size and geometry and the statistical properties of plastic deformation and energy dissipation in uniaxially compressed two-dimensional {\it small} crystals, by means of \add{atomistic simulations and analytical modeling.}


\section*{Methods}

We first consider the compression of perfect crystals of various sizes and aspect ratios. Crystals are simulated as two-dimensional aggregates of short-range interacting monodisperse particles in their lowest energy configuration, that is a triangular lattice in the $xy$ plane. Boundaries parallel to the $y$ direction are in contact with rigid walls, while those parallel to the $x$ direction are free (see Fig. 1). Uniaxial compression is applied symmetrically along the $x$ direction. We consider two different compression protocols: i) displacement-control, in which rigid walls are quasi-statically displaced at equal constant velocities in opposite directions; ii) force-control, in which the force exerted on the walls is slowly increased at a constant driving rate.
We consider a set of $N$ particles located at the nodes of a triangular lattice and confined between two rigid walls (as in Fig. 1).  Particles interact pairwise through a short-range potential $V_{ij}$. For simplicity, we use a Lennard-Jones potential, but any other short-range one would lead to qualitatively similar results. We implement overdamped dynamics simulations. This choice is inspired by experiments on colloids, in the presence of a viscous carried fluid.  The equation of motion for each particle $i$ at position ${\bf r}_i$

\begin{equation}
\Gamma \frac{d{\bf r}_i}{dt} = \sum_j {\bf f}\left(\frac{{\bf r}_i - {\bf r}_j}{d}\right)+{\bf f}_{d}({\bf r}_i),
\label{eq:vf}
\end{equation}
where $\Gamma$ is the viscous friction coefficient of the carrier fluid, $f$ is the interparticle force, $d$ the characteristic size of the particles, and $f_d$ is the corresponding driving force for each deformation protocol. We choose as units of space and time $d$ and $t_0=\Gamma d^2/V_0$ respectively, with $V_0$ the amplitude of the inter-particle potential, and we measure the driving force $f_d$ in units of $V_0/d$. \add{As a consequence we can take $V_0=1$ and $d=1$ without loss of generality.} In dimensionless units, the linear system sizes $L_x^0$ considered range from $L_x^0=16$ to $L_x^0=72$. Particles are confined by two rigid walls, that we model as two extra columns of particles of the same size, commensurate with the crystalline planes (see Fig. 1).  In displacement control simulations, rigid walls are driven quasi-statically  with a constant velocity $v=0.005$ in simulation units, whereas in the force control simulations the force exerted on the walls is slowly increased at a constant driving rate equal to $0.005$. In both cases, these values are sufficiently small to avoid the overlap of plastic events. Smaller driving rates do not yield significantly different results. \add{Upon compression, the strain $\gamma$ is measured as the ratio between the deformation along $x$ and the size of the undeformed sample, i.e. $\gamma=\Delta L_x/L_x^0 $ (engineering strain), while the applied stress $\sigma$ is the force per unit surface ($L_y^0 d$) exerted on the rigid walls.}    

The coupled Eqs.~(\ref{eq:vf}) for $i=1,...,N$ are integrated numerically with an adaptive step size fifth-order Runge-Kutta method with precision $10^{-6}$. Thermal and hydrodynamic effects on the particles have been neglected. Thermal fluctuations can be disregarded because the characteristic time for dislocation motion is much faster than the characteristic time for thermal diffusion. Moreover in the problem at hand, elastic interactions would prevail over hydrodynamic interactions between the particles, if in suspension.

\section*{Results and Discussion} 

\subsection*{Elastic loading and plastic yield}

In both protocols, the response is initially elastic. In a perfect crystal, the elastic limit is reached as soon as the motion of a pair of opposite sign edge dislocations is activated, as in Fig.  2, marking the beginning of the irreversible or plastic flow regime  (see Supporting Videos S1 and S2).  We use this limit to define the yield stress $\sigma_y$ and the yield strain $\gamma_y$ of the crystal. 

By moving, dislocations allow the system to slip plastically and emit/dissipate part of the stored elastic energy. The value of both the yield stress $\sigma_y$ and strain $\gamma_y$ are found to be independent of the deformation protocol. Figure 3 shows the dependence of the yield stress on the geometry of the sample under examination. Different systems sizes $L$ and aspect ratios $\rho=\add{L_x^0/L_y^0}$ are considered. Even below the microscale, we qualitatively recover the inverse size dependence of the yield stress that makes smaller samples stronger. 

According to the literature, in larger systems  the dependence of the yield stress on the system size follows a power law \cite{DIM-06}, which corroborates the view of plastic yield as a critical phenomenon, dominated by scale free rearrangements of dislocation patterns over finite-size effects \cite{Laurson-10}. Plasticity is dominated by bulk phenomena and surfaces are somehow peripheral to such mechanisms. At smaller scales, our results show instead that $\sigma_y$ is strongly affected by both size and geometry of the specimen. Size effects at this length scale were experimentally observed in compression tests in gold nanopillars \cite{GRE-06}, which were able to relate the increase in hardness to the {\it dislocation starvation} mechanism. Figure 2 shows that in our systems yield is mediated by a single dislocation pair (or even by a dislocation alone in the case of an imperfect crystal) and the onset of plasticity does not require the collective motion of a complex dislocation network. 
Dislocation interplay with boundaries thus becomes crucial. Early studies showed that the interaction of dislocations with rigid substrates is inherent to the strengthening of sheared thin films \cite{NIX-98,NIC-03}. Nevertheless, the role of rigid boundaries in uniaxial compression experiments is still a matter of investigation. 

The connection between the yield stress and the boundary effects can be easily visualized in our simple model system. Up to values of the stress very close to $\sigma_y$, the behavior of the system is elastic and the strain energy \add{per unit length} is approximated by 
\begin{equation}
E_e(\sigma_y)\approx \sigma_y^2\frac{ S}{2Y_\rho},
\end{equation}
 where $S$ is the specimen surface and $Y_\rho$ is the effective Young modulus. \add{Here and in the following we consider energies per unit length, given the quasi two-dimensional nature of the problem.}
At yield the onset of dislocation motion is almost instantaneous compared to the dynamics of steady loading. Dislocations must account for the stress distribution inside the compressed sample, giving rise to the elastic energy density $\mathcal{E}$. However, before dislocation motion is initiated, the elastic energy stored in the dislocated system  must be comparable to the energy  $E_{e}$ right before yield, as approximately no dissipation has occurred yet. This leads to the simple relation
\begin{equation}\label{energies}
E_e(\sigma_y) \approx\int\int_s \mathcal{E}dx\,dy.
\end{equation}
 Equation (\ref{energies}) establishes a connection between the yield stress and dislocation strain distribution. It also bears implicitly the information about the dependence of the yield stress on the system size and geometry. 
 
The essence of our problem then lies in the stress distribution that accounts for $\mathcal{E}$. In particular,  an anomalous stress concentration is required at rigid boundaries, in order to enforce the condition of vanishing displacements.  It is  evident that boundary conditions alter dramatically the energy landscape $\mathcal{E}$ in the specimen and ultimately affect the yield stress, as prescribed by Eq. (\ref{energies}).
 
\subsection*{An estimate of the yield stress} 
 
By means of elasticity theory, we can demonstrate that the stress fields produced inside the sample by an edge dislocation close to a rigid boundary are long ranged and decay as $\sim 1/r$, thus showing no screening effects, unlike dislocations close to free boundaries \cite{hirth}. For simplicity, we consider a positive straight edge dislocation with Burgers vector perpendicular to the rigid wall, a distance $l$ apart from the wall, as in Fig. 4 (left). The problem has translational invariance along the $z$ direction (plane strain conditions): the wall is at $x=0$, the dislocation is placed in $(l,0,z)$. We can calculate the exact displacement field ${\bf u}$ as $ {\bf u}={\bf u}^{inf}+{\bf u}^{img}+{\bf w}$ where ${\bf u}^{inf}$ is the displacement field of the original positive dislocation in $(l,0,z)$ in an infinite medium,
 ${\bf u}^{img}$ is the displacement of the image dislocation of opposite sign in $(-l,0,z)$ and ${\bf w}$ is and additional field, which is analytic in $x>0$ and satisfies the equilibrium elastic equations
 \begin{equation}\label{eq:1plane}
 \mu \nabla^2 {\bf w}+(\lambda+\mu)\nabla(\nabla \cdot{\bf w})=0,
 \end{equation}
 with boundary conditions such that the full ${\bf u}(0,y,z)=0$, being $\mu$ the shear modulus and $\nu$ the Poisson ratio. Eq. (\ref{eq:1plane}) with the given boundary conditions is commonly known as the $1-$plane problem in linear elasticity. Several techniques can be devised to find exact solutions \cite{love} and the complexity resides solely on the complicated form of boundary conditions.  Here we follow \cite{tedone} as reviewed by \cite{love} and observe that Eq. (\ref{eq:1plane}) can be rewritten as a Laplace equation in the form
\begin{eqnarray}\label{eq:laplace}
\nabla^2\left[ {\bf w} + \frac{\lambda+\mu}{2\mu}\left(
\begin{array}{c}
x\\y\\z\\
\end{array}
\right)
\vartheta
\right]=0,
\end{eqnarray}
where $\vartheta=\nabla\cdot {\bf w}$ is the dilatation. The problem can then be solved using the Green function method, being the Green function for  Eq. (\ref{eq:laplace}) in the $1-$plane geometry known from classical electrostatics. While further details will be given in a future publication \cite{MOR-11}, for the problem at hand we obtain    
\begin{eqnarray}\label{eq:solution}
w_x&=&  \frac{1}{\pi} \left[\arctan\frac{y}{l+x}+\frac{1}{2(1-\nu)}\frac{ly}{(l+x)^2+y^2}\right]+ \nonumber\\
&&+ \frac{x}{\pi(3-4\nu)}\left[\frac{y}{(l+x)^2+y^2}+\frac{l}{2(1-\nu)}\frac{2y(l+x)}{[(l+x)^2+y^2]^2}\right] \\
w_y&=& -\frac{x}{\pi(3-4\nu)}\left[\frac{l+x}{(l+x)^2+y^2}+\frac{l}{2(1-\nu)}\frac{(l+x)^2-y^2}{[(l+x)^2+y^2]^2}\right] \nonumber\\
w_z&=&0. \nonumber
\end{eqnarray}

From the above result, the \add{elastic} strain \add{tensor} $\varepsilon_{ij}$, the stress  tensor  $\sigma_{ij}$ and the elastic strain energy density 
\add{$\mathcal{E}$} can be calculated using linear elasticity. Interestingly,
while the stress components derived from ${\bf u}^{inf}+{\bf u}^{img}$ decay like $\sim 1/r^2$ (being that the stress of a dislocation dipole), the stress due to ${\bf w}$ will decrease like $\sim 1/r$. Overall, the stress field of an edge dislocation close to a rigid boundary will hence be long-ranged, as opposed to the case a free boundary. As a consequence,
stress fields are able to sample the entire system size and are responsible for the size-sensitivity of the yield stress. 

In the case of our simulations, however, Eq. (\ref{eq:solution}) may seem of limited help in calculating the strain energy  $\mathcal{E}$ at yield for our model system, where we have two fixed boundaries, two free surfaces and two dislocations of opposite signs, as the exact solution of such a problem would be great complexity. Nevertheless, an approximation of the integral in Eq. (\ref{energies}) can be provided as follows. Fig. 4 (right) shows the result of the superposition of the solutions for the shear stress of two edge dislocations of opposite signs, as calculated from Eq. (\ref{eq:solution}), arranged in a configuration which mimics the one observed at yield in our simulations for the perfect crystal. For simplicity we consider Burgers vectors along $x$, but the general solution would lead to similar results \cite{MOR-11}. In such a configuration, the condition of zero displacement at the vertical walls is not met anymore, however one can prove that the deviations from zero affect the stress and the strain energy in a negligible way if the dislocations are far apart. At the same time, Fig. 4 (right) shows that the two dislocations behave like a dipole, in the sense that the stress goes rapidly to zero outside the region enclosed by them, approximating the stress field close to free boundaries. Such observation is verified analytically and is valid for all stress components.   

We can conclude that  the configuration in Fig. 4 (right) provides an acceptable approximation for the elastic problem in our simulations and an estimate for the strain energy 
$\mathcal{E}$ in Eq. (\ref{energies}).  Finally, upon inverting Eq. (\ref{energies}) and computing numerically the integrals of $\mathcal{E}$, we evaluate the yield stress and its dependence on the system size and geometry. \add{Predictions from our theory} are given in Fig. 3 \add{(solid lines)} and are in remarkably good agreement with \add{our}  simulation results, \add{presented below} \add{(discrete points)}, suggesting that boundaries and shape effects are essential in our problem. Given the nature of the approximations involved, we stress that ours is just a dimensional estimate. However the agreement with simulations  corroborates  the view that  the role played by boundaries is integral to nanoplasticity.

\subsection*{Plastic avalanches}

As soon as the yield point is reached, the response of the system to further loading differentiates depending on the deformation protocol. Under conditions of displacement control, stress-strain curves are characterized by serrated yielding, while they assume a staircase shape under conditions of stress control. We emulate realistic realizations of compressed samples by introducing randomness at free boundaries \add{as follows. The initial state of each realization is obtained from the perfect crystal by extracting a random number of particles from one free surface and relocating them at random positions on the opposing surface. In this way both the number of particles and the linear size of all simulated specimens are kept constant, while the morphologies of their free surfaces are allowed to vary stochastically.}  
Strain plateaus $\Delta \gamma$ in force control or stress drops $\Delta \sigma$ in displacement control always correspond to plastic events in which dislocation motion is reactivated in order to reduce the elastic energy stored during prior loading phases (see Fig. 1). Remarkably, a statistical analysis of event sizes over several realizations of surface disorder (Fig. 5) reveals that sizes are power-law distributed as 
\begin{equation}
p(\Delta\sigma)\sim \Delta\sigma^{-\tau_\sigma}
\;\;\;
p(\Delta\gamma)\sim \Delta\gamma^{-\tau_\gamma}.
\end{equation} 

Due to the limited system size, moving dislocations easily leave the sample through free boundaries. Pioneering studies have shown that in sub-micrometer Ni samples, pure mechanical loading can induce dislocation depletion within the sample \cite{SHA-08}. In our case plastic flow proceeds through the activation  and motion of few dislocations at a time, while no storage is observed (see Supporting Video S3). Such phenomenology differs from the widely accepted picture of plastic flow at larger scales. Yet, our statistical study leads to broadly distributed plastic events for several system sizes, suggesting that kinematic constraints and long range dislocation interactions still rule plastic flow, as shown in Fig.   5. The cumulative distribution is defined as $p_c(x)=\int_0^x p(x')dx'$. Upon a closer inspection, a noticeable difference in the exponent value is nevertheless encountered, \add{as we show in the following}.

Plastic event sizes are commonly quantified in experiments by looking at the amount of energy $W_{dis}$ released during each event or, under stress control conditions, by recording the magnitude of platen displacements. 
\add{
By defining the stored potential energy as $U=\sum_{i<j}V_{ij}$, each variation of $U$ in our simulations will be given by the energy balance relation
\begin{equation}\label{eq:balance}
\Delta U=W_{dis}+W_{ext}.
\end{equation}
Here the dissipated energy  $W_{dis}$ can be measured as the time integral of the dissipated power, i.e. $W_{dis}\equiv \int dW_{dis}=\sum_i\int f_i v_i dt$ (force time velocity for all particles), whereas $W_{ext}$ is the work done by the external load, $W_{ext}=\int \sigma d\gamma$. The strain is the sum of its elastic and plastic components $\gamma=\gamma_{el}+\gamma_{p}$.
As in our system no defect accumulation is encountered, we have $\Delta \sigma \propto \Delta \gamma_{el}$ (linear elasticity) and evidently $\Delta U\sim (\Delta \gamma_{el})^2$.  

Let us first consider the case of force control. If focusing on a single platen displacement event, we have $\Delta \sigma \simeq 0$ and $\Delta \gamma_{el}\simeq 0$. This implies that the change in stored elastic energy is roughly zero, $\Delta U\simeq 0$, and, according to Eq. \ref{eq:balance}, $W_{dis}\simeq W_{ext}$, which means that the dissipated energy comes essentially from the work done during the platen displacement event of magnitude $\Delta \gamma\simeq \Delta \gamma_{p}$.  
}
Then evidently $W_{dis}\sim \Delta \gamma_p$, as the amount of slip due to a moving dislocation is proportional to the energy dissipated in the process \cite{ZAI-05}. This relation indeed \add{is verified in our simulations}, as shown in Fig. 6(a).
Experimental studies at larger scales show that both quantities follow a power-law distribution, decaying with an exponent close to the mean-field value $\tau=3/2$  \cite{ZAI-05}. The striking aspect of the exponent $\tau$ is its universality \cite{CSI-07}. In our model, which aims to reproduce nanoscale plastic flow, the exponent $\tau_\gamma \sim 0.9$ deviates substantially from the universal value, suggesting that at such scales  the microscopic dislocation dynamics giving rise to plastic deformation are qualitatively different. Signals of universality breakdown also come from experiments of compressed high-purity LiF micropillars  \cite{DIM-09}. 

Under displacement control conditions, instead, the statistical analysis of plastic flow and dissipated energy can be performed by looking at the distribution of stress drops (Fig. 5(b)). 
In Fig. 6(b) we show the evolution of the energy balance during compression. For quasi-static driving,  stress drops occur almost instantaneously, i.e. at constant total strain 
($\Delta \gamma \simeq 0$, or $W_{ext}\simeq 0$). This implies that, during the stress drop, a sudden decrease of the elastic strain $\Delta \gamma_{el}$ of the crystal must occur at the expense of 
increasing its plastic strain, $|\Delta \gamma_{el}| \simeq |\Delta \gamma_{p}|$.  
\add{According to Eq. \ref{eq:balance}, being $W_{ext}\simeq 0$ during the stress drop,  the energy 
dissipated by plastic  avalanches must balance the change of internal energy, $\Delta W_{dis}\simeq \Delta U$. Then, as in the the case of force control, since there is no accumulation of 
large numbers of defects, $\Delta \sigma \propto \Delta \gamma_{el}$ and $\Delta U \sim (\Delta \sigma)^2$
}.
Being stress drops distributed according to a power-law of exponent $\tau_\sigma$, we obtain the following exponent relation $\tau_E=(\tau_{\sigma}
+1)/2$ for the dissipated energy distribution, which for $\tau_{\sigma} \sim 0.7$ yields $\tau_E \sim 0.85$, in good agreement with the results obtained for $\tau_{\gamma}$ in the 
force control protocol.   
\add{We emphasize that stress drops in displacement-controlled deformation are in principle different from stress increments between bursts in force-controlled tests. Recent experimental studies have shown the stress increments are Weibull-distributed in Mo micro- and nanopillars \cite{ZAI-08}. 
} 

\add{
Compared to numerical studies of size effects in dislocation dynamics at the microscale \cite{CSI-07}, which highlight a strong size dependence of the avalanche distribution cutoff, our small systems span a very narrow linear size range. Plasticity is thus always mediated by equally small numbers of dislocations and maximum avalanche sizes for systems with initial size $L_x^0=16$, $24$ and $32$ do not give rise to appreciable differences in the avalanche-size cutoff as plastic flow advances. It should also be noted that by definition of $\gamma$, platen displacements are always $\Delta \gamma <1$, and similarly under displacement control, large stress drops are bounded by the very nature of the short-range interatomic potential considered.
}

\add{
As for the origin of the anomalous avalanche exponents, we should remark that the novel behavior is related to the inability of the system to store large numbers of dislocations. In the absence of collective behavior, plastic flow departs from the traditional picture of cooperative dislocation organization. In fact our simulations show a behavior which approximately recalls a sequence of load-unload events, much in the spirit of stick-slip dynamics or fracture/failure mechanics. We notice that an energy release exponent very close to $1$ has been recently encountered in simulations of yielding and failure of heterogeneous materials in the plastic steady state \cite{PIC-10}. However the rationale behind the analogy with our finding remains to be ascertained. 
Finally, a statistical analysis of surface roughness of strained systems would provide further insights into the nature of plasticity at such scale. However, linear sizes of our free surfaces are only one order of magnitude larger than the lattice constant and such a study would not lead to significant results. Further studies in this direction are currently under way.     
}

\subsection*{Conclusions}

In conclusion, we have shown that the onset of plasticity at small scales is mediated by few dislocations. The number and arrangement of nucleated dislocations must account for the distribution of stress stored inside the crystal during the elastic-loading regime, allowing one to estimate the dependence of the yield stress on  sample size and geometry.  Our  results confirm that both size and shape are crucial factors in determining the strength of materials at these scales. We find that plastic flow occurs in an intermittent manner reminiscent of irreversible deformation at larger length scales. Plastic avalanches of broadly distributed sizes are still observed,  however, the absence of dislocation storage has important effects on the scaling characteristics of viscoplastic dynamics, which ultimately violate the {\it universal} mean-field behavior observed at larger scales. 
\add{Our results are thus a significant example of {\it source-limited deformation} \cite{SHA-08} and arise naturally from the impossibility of such a small system to store dislocations}. The new exponent values are obtained for a two-dimensional crystal geometry and  should be relevant for thin films of several self-assembled nanoparticles under external loading conditions and amenable of experimental analysis in colloidal systems.

\section*{Acknowledgments}
The Authors are grateful to M. Zaiser, L. Laurson and I. Groma for stimulating discussions.


\section*{Figures}
\begin{figure}[!ht]
\centering 
\epsfig{file=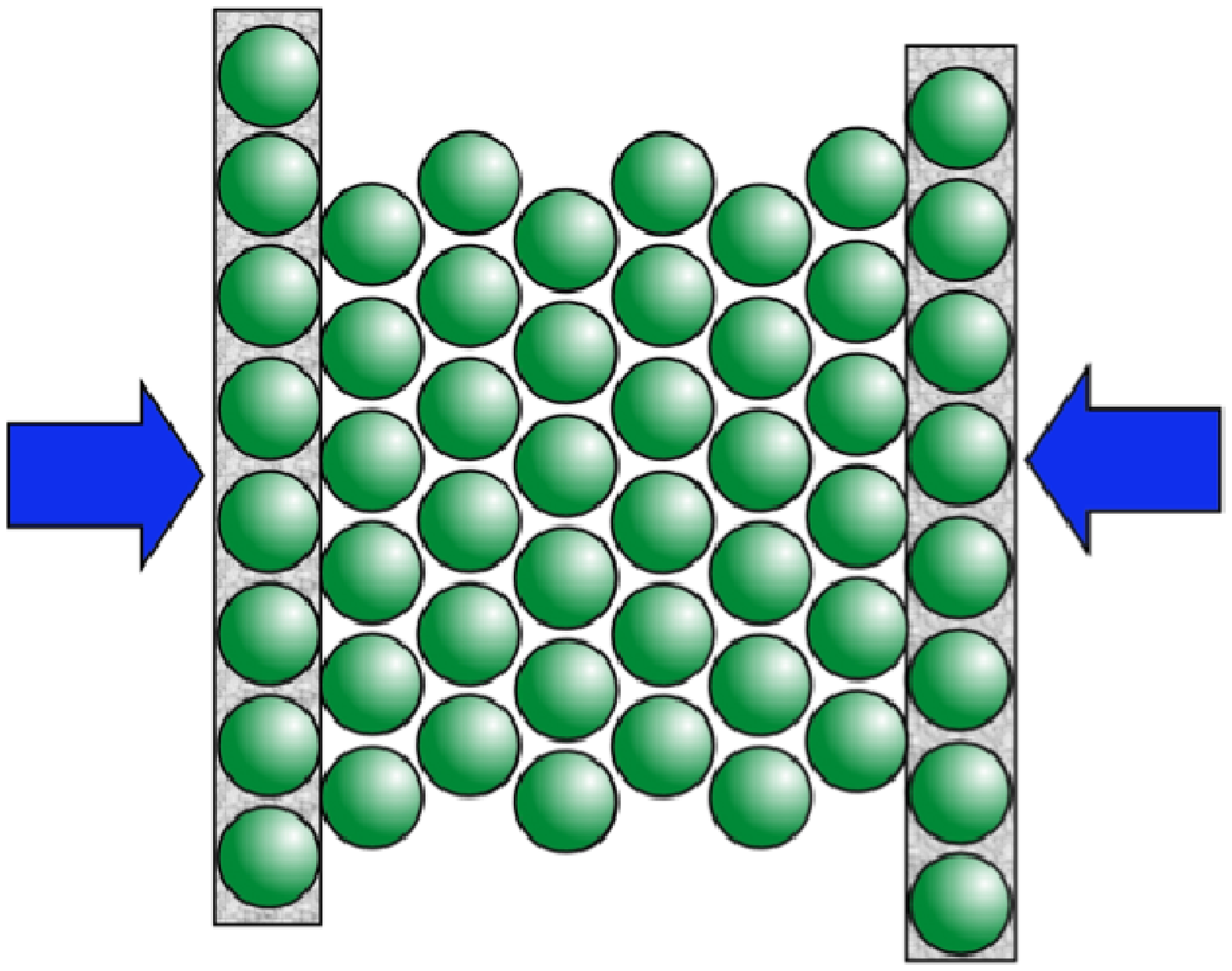,width=6.5cm,clip=}\epsfig{file=stress-strain-control.eps,width=7.5cm,clip=}\caption{
{\bf System and stress strain curves.}   (Left) Schematic representation of a uniaxial 
compression test on a perfect crystal monolayer. (Right) We obtain i) a serrated-flow stress response 
under displacement (or strain) control conditions, i.e. ruling the position of the 
bounding walls (inside the grey boxes), or ii) a staircase shaped curve under force
 or stress control conditions, i.e. regulating the applied force on the driving walls. The dashed line 
signals the location of the yield point.
}
\end{figure}

\begin{figure}[!ht]
\centering  
\hspace*{0cm}\epsfig{file=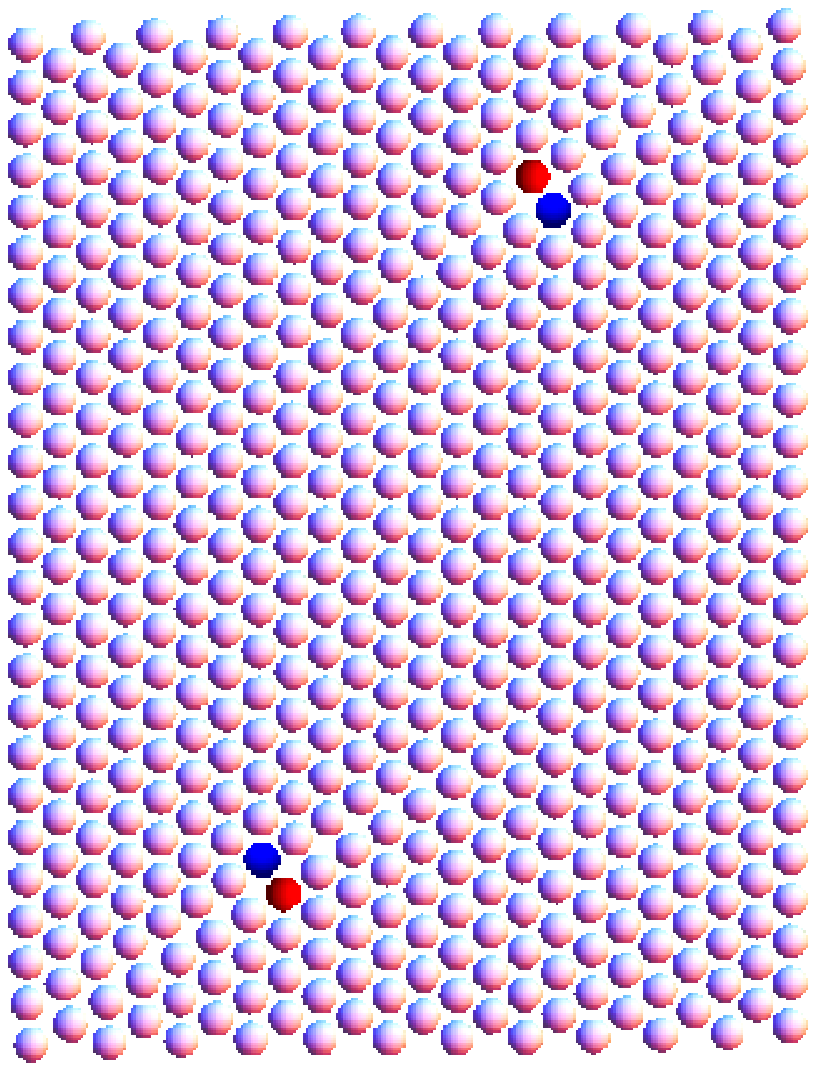,height=6cm,clip=}\hspace*{-1cm}\epsfig{file=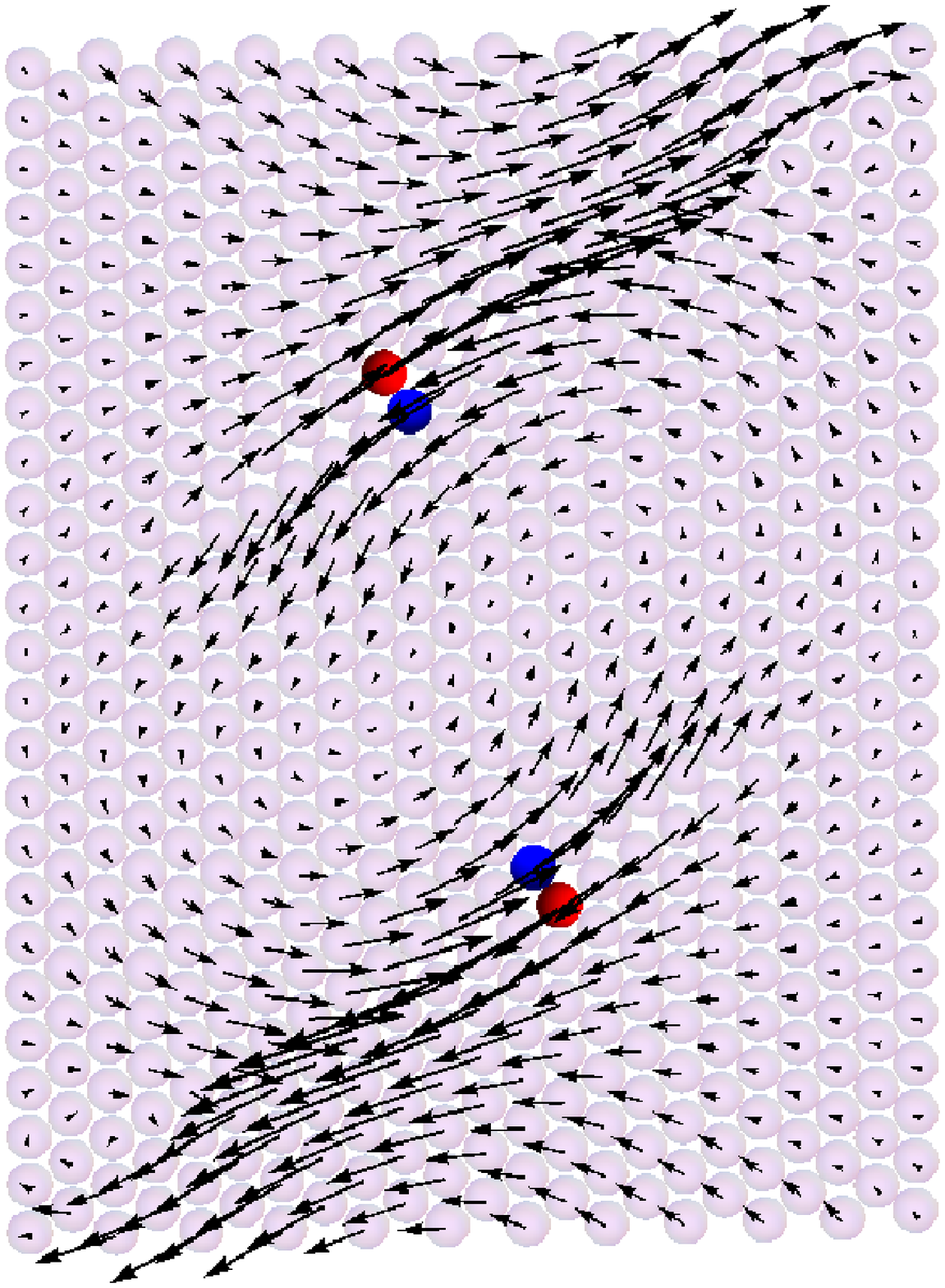,height=6cm,clip=}\hspace*{-1cm}\epsfig{file=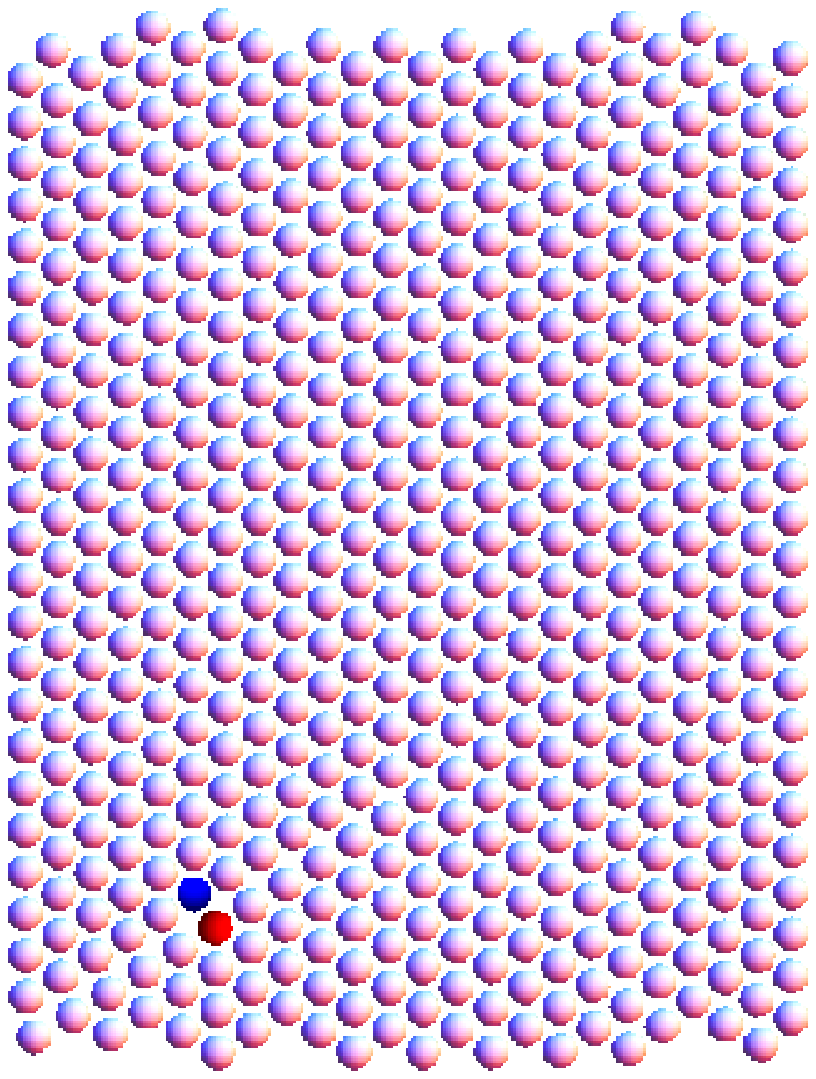,height=6cm,clip=}
\caption{
{\bf Dislocation nucleation at yield.}  
(Left) Perfect crystal, a dislocation pair is nucleated. 
Dislocations are represented as pairs of 5- and 7-coordinated particles, in blue and red respectively. 
(Center)  Instantaneous velocity field of the particles.  (Right) Surface-disordered crystal, dislocations may 
be individually nucleated at surface steps. 
}
\end{figure}

\begin{figure}[!ht]
\centering \epsfig{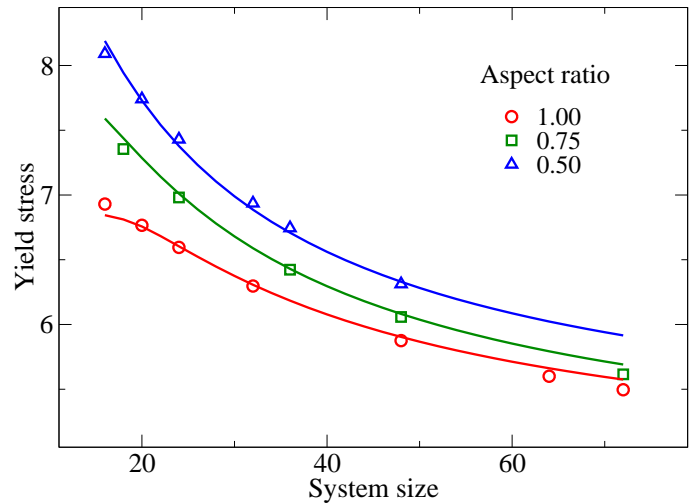}
\caption{
{\bf Yield stress in our nano-scale geometry.}  
 Smaller systems are stronger,
 but also different aspect ratios $\rho$ induce different yield patterns. Discrete data: simulation results. Solid lines: theoretical predictions (see main text).}
\end{figure}

\begin{figure}[!ht]

\centering 
\epsfig{file=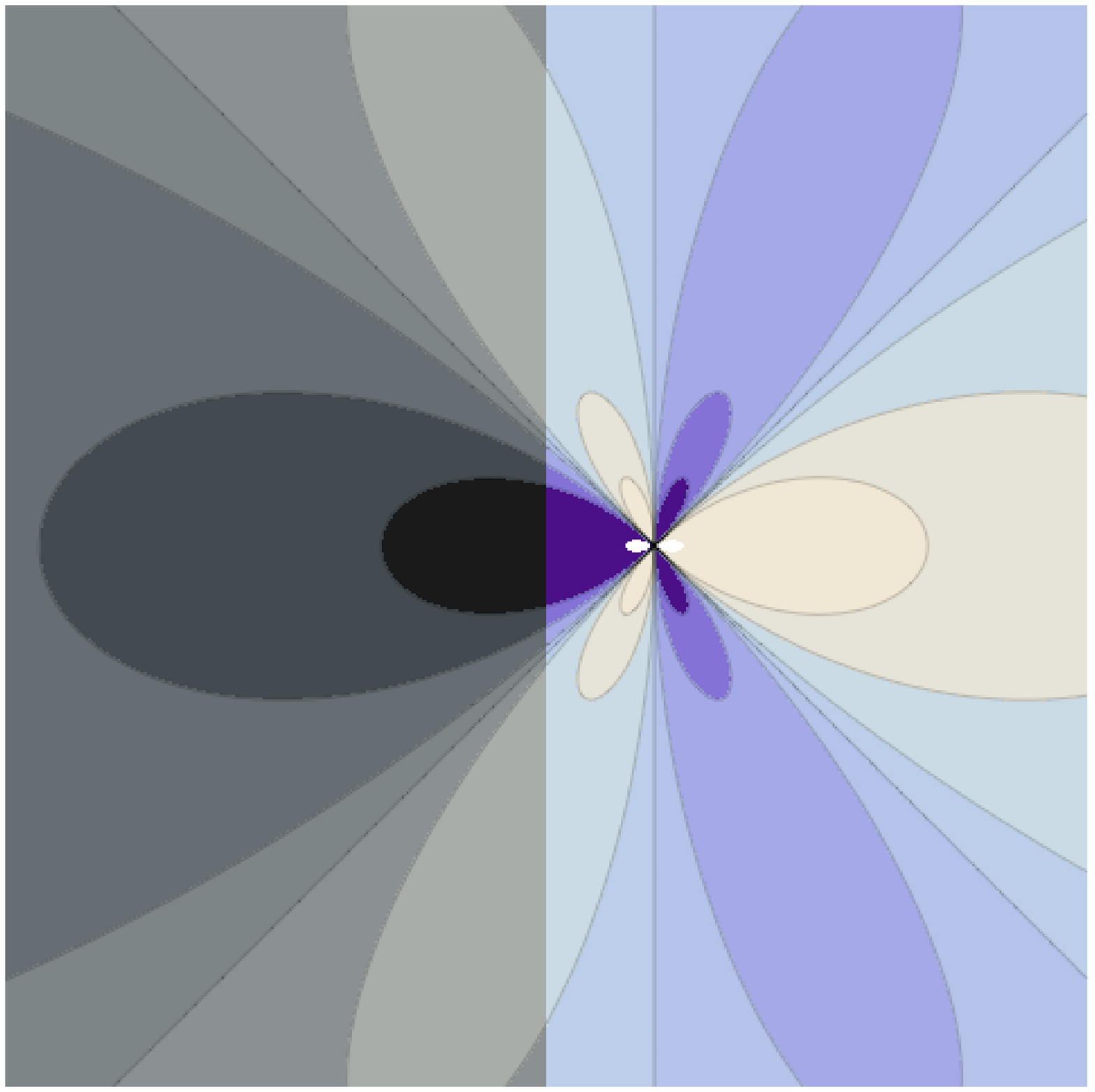,height=6.5cm,clip=}\hspace*{0.1cm}\epsfig{file=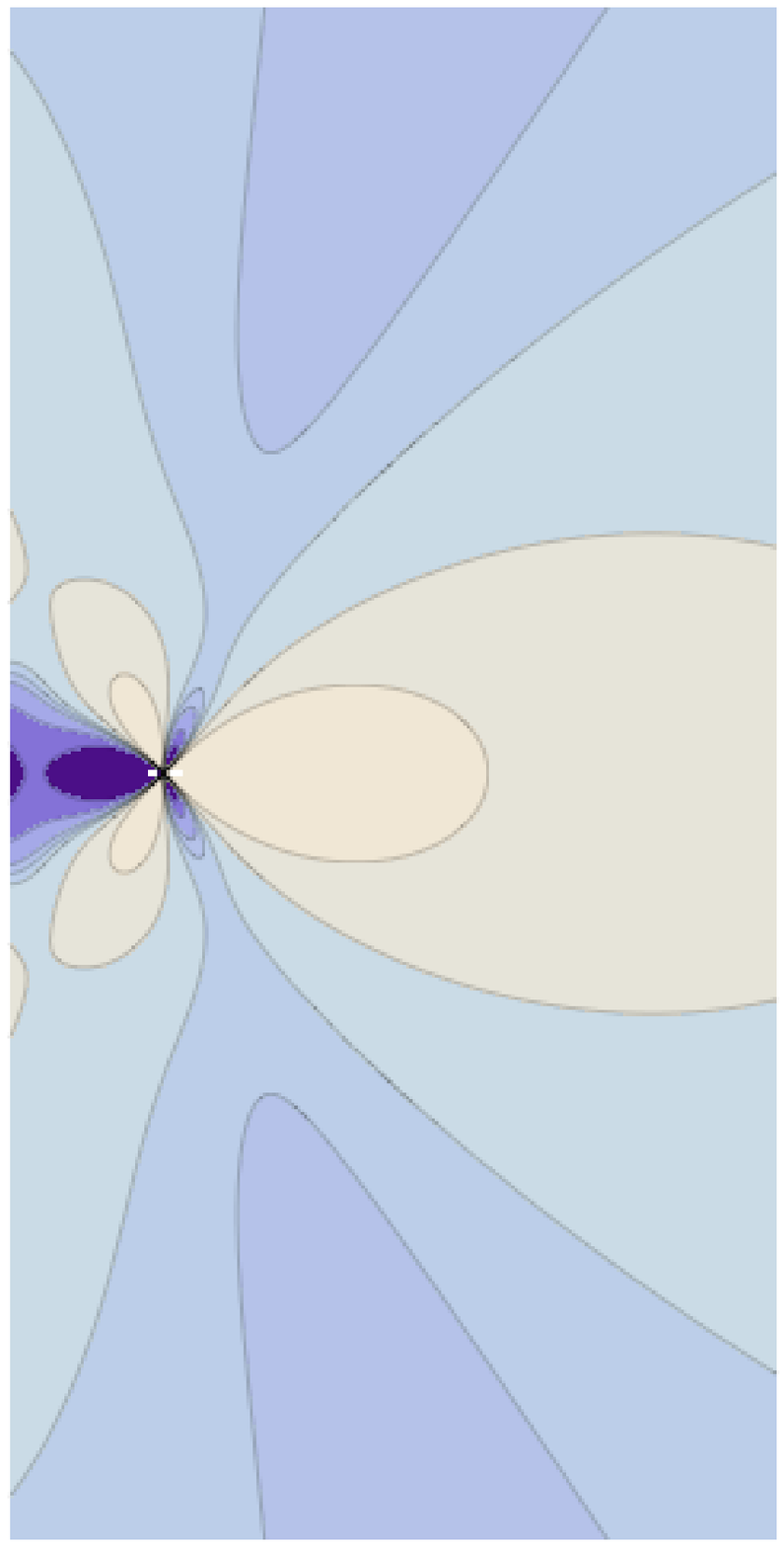,height=6.5cm,clip=}\hspace*{0.1cm}\epsfig{file=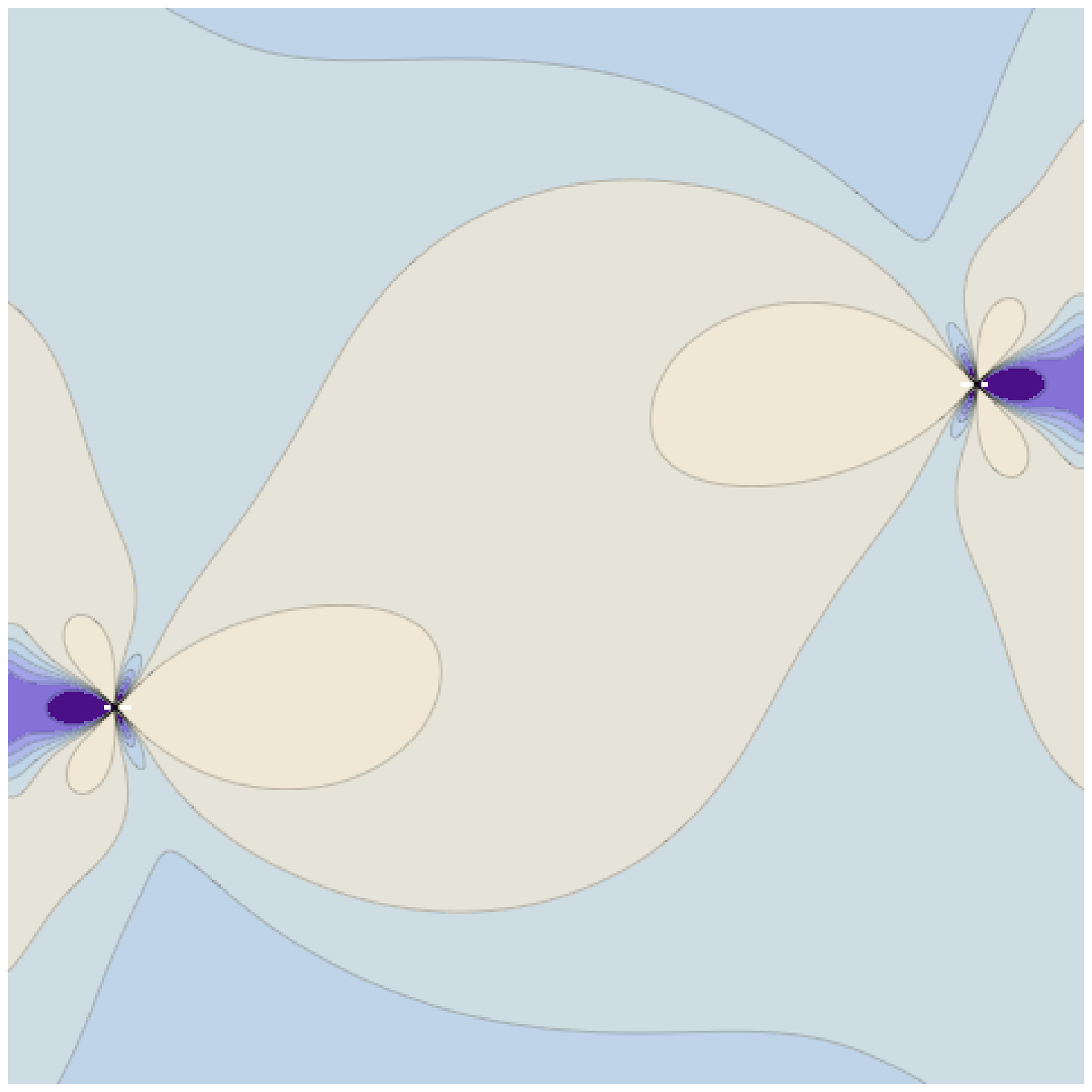,height=6.5cm,clip=}
\caption{
{\bf The role of rigid boundaries.}  (Left) Shear stress distribution around an edge dislocation in an infinite medium. 
The right side of the picture is shaded, indicating that we are interested in how the
 stress field changes upon imposing a rigid boundary. (Center) Shear  stress distribution around an 
edge dislocation located near a rigid boundary. (Right) Shear stress distribution due to a pair of 
opposite sign dislocations, confined within two rigid walls. Stresses rapidly vanish in the $y$ direction.  
Color code: Lighter regions indicate positive stresses, darker regions negative stresses. 
Vanishing stresses are represented in light blue.
Curves of equal stress are represented as a guide to the eye. For simplicity, the case of Burgers vectors perpendicular to the walls was considered here.}
\end{figure}

\begin{figure}[!ht]
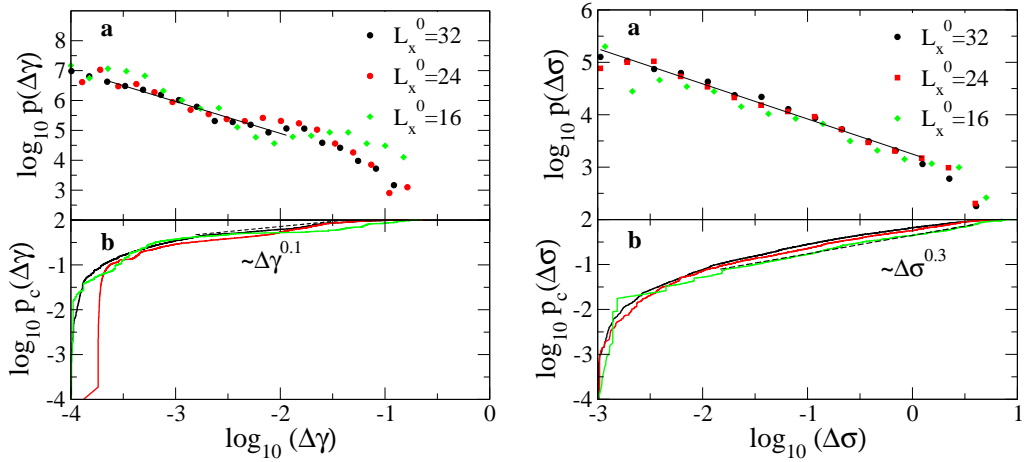

\centering 
\epsfig{file=avalanches-protocoloII.eps,width=6.5cm,clip=}\hspace{0.5cm}\epsfig{file=avalanches-protocoloI.eps,width=6.5cm,clip=}
\caption{
{\bf  Statistics of plastic events.} Left: (a) Distribution of platen displacements $p(\Delta \gamma)$
 in force-control simulations, decaying with an exponent $\tau_\gamma\approx 0.9$.  
(b) Cumulative distribution $p_c$ of platen displacements. 
Right: (a) Distribution of stress drops in displacement-control simulations, 
decaying with an exponent $\tau_\sigma\approx 0.7$. (b) Cumulative distribution of stress drops $p_c(\Delta \sigma)$
}
\end{figure}

\begin{figure}[!ht]
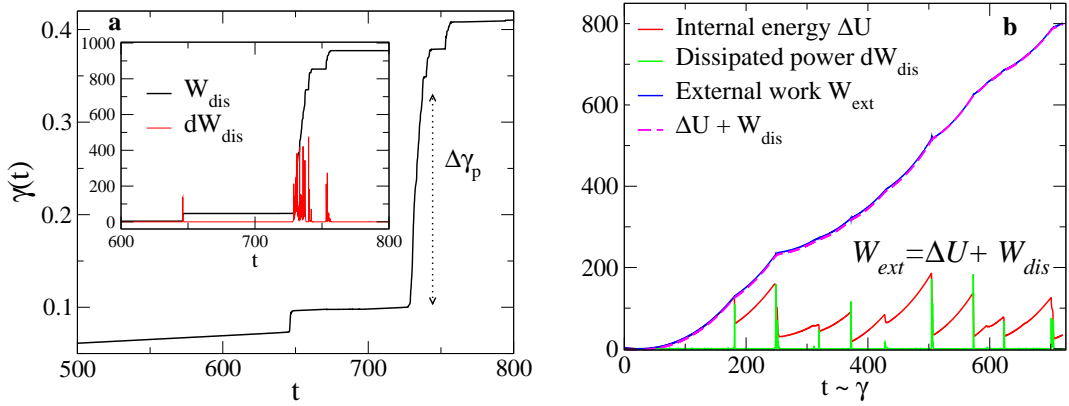

\centering
\epsfig{file=strain-energy-v3.eps,width=7cm,clip=} \hspace{0.5cm}\epsfig{file=24x24-first-principle.eps,width=6.5cm,clip=}
\caption{
{\bf Energy balance during plastic flow.}  (a) Temporal evolution of the nominal strain $\gamma$ and (inset) the 
dissipated energy in a force control compression test. In the quasi-stationary limit, 
platen displacements occur at constant stress and the dissipated energy is proportional 
to the increase of plastic strain. (b) Energy balance in a displacement control compression test.
 Stress drops occur at almost constant strain.
}
\end{figure}

\section*{Supporting Videos}
\begin{itemize}
\item{
{\bf Video S1} 
The first appearance of a dislocation pair signals the onset of yield for the perfect system. As soon as each dislocation reaches the opposite rigid boundary, plastic activity stops and the the first event is over. The animation consists of 7 snapshots of the dynamics. Top: dislocations are represented as pairs of 5- and 7-coordinated particles, in blue and red respectively. Bottom: velocity field corresponding to the dislocation configuration above. The modulus of the velocity vector is represented. Lower velocities are in red, higher in violet, according to the color scheme of the visible spectrum.
}
\item{
{\bf Video S2}
Time evolution of the velocity field at the time-steps shown in Video 1. Moving dislocations trigger particle motion and thus elastic energy dissipation. 
}
\item{
{\bf Video S3} 
Dislocation dynamics in the flow regime. Plastic events correspond to the activation of few dislocations at a time. Dislocation storage is not observed for such small systems. The animation consists of 20 snapshots of the dynamics. Conventions are as in Video 1.
}
\end{itemize}


\end{document}